# Confinement of magnetism in atomically-thin $La_{0.7}Sr_{0.3}CrO_3$/$La_{0.7}Sr_{0.3}MnO_3$ heterostructures.

**Authors**

Sanaz Koohfar[1], Alexandru Bogdan Georgescu[2], Aubrey Penn[3], James M. LeBeau[3], Elke Arenholz[4], Divine Philip Kumah[1*]

**Affiliations**

[1]Department of Physics, North Carolina State University, Raleigh, NC, 27695, USA.

[2]Center for Computational Quantum Physics, Flatiron Institute, 162 5th Avenue, New York, NY 10010, USA.

[3]Department of Materials Science and Engineering, North Carolina State University, Raleigh, NC, 27695, USA.

[4]Advanced Light Source, Berkeley, CA, 94720, USA.

**Abstract**

At crystalline interfaces where a valence mismatch exists, electronic and structural interactions may occur to relieve the polar mismatch leading to the stabilization of non-bulklike phases. We show that spontaneous reconstructions at polar $La_{0.7}Sr_{0.3}MnO_3$ interfaces are correlated with suppressed ferromagnetism for film thicknesses on the order of a unit cell. We investigate the structural and magnetic properties of valence-matched $La_{0.7}Sr_{0.3}CrO_3$ - $La_{0.7}Sr_{0.3}MnO_3$ interfaces using a combination of high-resolution electron microscopy, first principles theory, synchrotron X-ray scattering and magnetic spectroscopy and temperature-dependent magnetometry. A combination of an antiferromagnetic coupling between the $La_{0.7}Sr_{0.3}CrO_3$ and $La_{0.7}Sr_{0.3}MnO_3$ layers and a suppression of interfacial polar distortions are found to result in robust long range ferromagnetic ordering for ultra-thin $La_{0.7}Sr_{0.3}MnO_3$. These results underscore the critical importance of interfacial structural and magnetic interactions in the design of devices based on two-dimensional oxide magnetic systems.

*dkumah@ncsu.edu

**MAIN TEXT**

**Introduction**

Structural, electronic and magnetic interactions at the interfaces between thin films of crystalline polar transition metal perovskites have led to the realization of a wide range of physical phenomena not found in the bulk constituent materials. These exotic phenomena, which include, two-dimensional electronic gases, interfacial magnetism and superconductivity and orbital ordering are driven by interfacial chemical, structural, electronic and orbital reconstructions which serve to alleviate the polar mismatch at these interfaces.[1, 2] A consequence of these reconstructions is that significant deviations in the atomic scale structural and electronic properties of atomic layers adjacent to the polar interfaces arise. Due to the strong coupling of the structural, electronic and magnetic properties in these materials, detrimental effects may be induced in layers close the interface leading to strong thickness-dependent physical properties.[3-5] To confine the functional properties of transition metal perovskites to two-dimensions in order to realize novel phenomena associated with quantum confinement and engineer novel device architectures including spin-tunnel junctions, understanding and controlling these interfacial interactions is crucial. Additionally, in multiferrroic heterostructures where the magnetoelectric coupling effect is confined to interfacial layers,[6] reducing the thicknesses of the component layers to the order of a unit cell will lead to an enhanced coupling of ferroic order parameters.

An important system which displays thickness-dependent transitions related to interfacial interactions is the $LaSrMnO_3$ (LSMO)/$SrTiO_3$ (STO) interface. LSMO films have been explored for their half-metallic properties and colossal magnetoresistance effects with applications in spintronic devices.[6] The magnetic and electronic properties are related to the Mn-O bond properties in neighboring unit cells through double exchange interactions and Jahn-Teller effects.[7] Distortions to the Mn-O bond angle achieved by chemical substitution and strain results in a modulation of the magnetic and electronic phases.[8] When grown as epitaxial thin films, properties of LSMO can be coupled to other functional oxides including ferroelectric and superconducting perovskite transition metal oxides.[6, 9, 10]

LSMO films have been reported to undergo a ferromagnetic to paramagnetic transition for film thicknesses below 4-10 unit cells (uc) when grown on lattice-matched STO substrates.[3, 11] The thickness-dependent phase transition has been attributed to interfacial interactions driven by the polar mismatch between the two materials. These interactions include interfacial ionic intermixing which leads to deviations in the composition of interfacial LSMO layers,[12, 13] interfacial charge transfer evidenced by X-ray absorption spectroscopy and electron microscopy measurements, and ferro-distortive ionic displacements.[14-16]

To eliminate the interfacial interactions which lead to suppressed magnetism, we show that the insertion of $La_{1-x}Sr_xCrO_3$ (LSCO) spacer layers at LSMO interfaces removes polar structural distortions observed at LSMO interfaces[14] and couples to the lattice symmetry of LSMO leading to robust ferromagnetism down to 2 uc in LSCO (M uc)/LSMO (N uc)/LSCO (M uc) (M/N/M) heterostructures grown on (001)-oriented STO substrates by molecular beam epitaxy. The role of the LSCO spacer is two-fold: (1) By matching the La/Sr ratio of the LSCO to the LSMO film, the polar mismatch at the LSMO interface is effectively removed and (2) LSCO which has a $R\bar{3}c$ symmetry with $aa^-a^-$ rotations (160º Cr-O-Cr bond angle, 3.88 Å pseudocubic lattice constant)[17] couples to the oxygen octahedral rotations in LSMO alleviating the oxygen-octahedral mismatch.[18, 19] Additionally, we observe an antiferromagnetic exchange interaction between the Cr and Mn ions at the LSMO/LSCO interface. Using a combination of temperature-dependent magnetometry, picometer-scale synchrotron X-ray based /structural characterization and element-specific magnetic spectroscopy, we demonstrate the enhancement of ferromagnetic ordering in ultra-thin LSMO films, and the stabilization of bulklike ferromagnetism in LSMO layers as thin as 2 uc (0.8 nm). The results are confirmed by first principles theory.

**Results**

LSMO films and LSCO (M)/LSMO (N)/LSCO (M) (M/N/M) heterostructures and [LSCO (M)/ LSMO (N)] superlattices were synthesized by plasma-assisted oxide molecular beam epitaxy at a growth temperature of 800 °C. The LSMO thickness, N, was varied from 2 to 10 uc while the LSCO thickness, M, was fixed at either 2 or 3 uc to be above the decay length for surface polar distortions observed for LSMO[14] and LaNiO$_3$ [20] films. A schematic of the heterostructures is shown in Figure 1(a). As shown in Figure 1(b), the LSCO layers possess the same +0.7/-0.7 net charge stacking along the growth direction. Hence, no polar discontinuity exists at the LSMO top and bottom interfaces. RHEED oscillations are observed for all the layers indicative of layer-by-layer growth as shown in Figure 1(c). The surface morphology of the films were characterized by atomic force microscopy. A representative atomic force microscope image is shown in Figure 1(d) for the 3/3/3 sample with a surface roughness of less than 1 unit cell.

*High-resolution electron microscopy measurements*

The aberration-corrected high angle annular dark field-scanning transmission electron microscopy (HAADF-STEM) (Z-contrast) image in Figure 1(e) shows the 2/6/2 heterostructure. Energy-dispersive x-ray spectroscopy (EDS) chemical maps reveal some chemical intermixing within a unit cell between Mn and Cr layers and Ti migration across the LSCO/STO interface.[21] In contrast, the La signal drops abruptly at the interface. Electron energy-loss spectroscopy (EELS) was also performed in the STEM to track relative variations in Mn and Cr oxidation state across the film with the $L_3/L_2$ white line ratio according to the method described by Tan, *et al*.[22]

*Magnetization Measurements*

The magnetic properties of the LSMO thin films and M/N/M heterostructures and [M/N] superlattices were characterized using a Quantum Design SQUID system. The magnetization curves normalized to the Mn are shown as a function of field and temperature at 10 K for M=3 and N=2,3,4,6 and 10 in Figure 2(a) and 2(b), respectively. Temperature-dependent curves were recorded after field cooling the samples in a 1 Tesla field applied in-plane. As will be discussed later, the magnetization in the LSMO sublattice is determined to be closed to bulk, and the reduced net SQUID magnetization as the LSMO thickness decreases is related to the contribution of anti-parallel spins in the LSCO layers resulting from antiferromagnetic interactions between the LSMO and LSCO layers.

In contrast to the single layer LSMO films, the 3/N/3 heterostructures remain ferromagnetic down to N=2. Inserting 3 uc LSCO at the LSMO top and bottom interfaces results in ferromagnetic ordering below~150 K compared with the 2 uc LSMO films grown directly on STO which remains paramagnetic down to 2 K. To verify that the magnetism in the heterostructures is not within the LSCO layers alone, the magnetization for a 6 uc LSCO layer was measured and found to be similar to the STO substrate down to 2 K as shown in Figure S1 of the supplementary materials.[23] The results for the LSCO film are consistent with bulk LSCO which is a G-type antiferromagnet.[17]

We further elucidate the magnetic coupling at LSCO/LSMO interface by measuring the magnetization hysteresis loops after cooling down the sample in ±0.5 T applied in-plane field. Figure 2(c) demonstrates the presence of an exchange bias for a [LSCO (2)/ LSMO(4)]x6 superlattice. There is a negative (positive) shift in the hysteresis loop when a positive (negative) field has been applied, indicating reversible switching of exchange bias. The exchange bias field is 332 Oe for both positive and negative field-cooled hysteresis loops.[24, 25]

*X-ray Circular Dichroism Measurements*

To further confirm that the observed ferromagnetism in the 3/N/3 heterostructures occurs in the LSMO layers, we performed element specific X-ray magnetic circular dichroism (XMCD) measurements at the Mn and Cr $L_{2,3}$ absorption edges at 4.0.2 beamline at the Advanced Light Source using total electron yield. The absorption spectra at the Cr and Mn edges are consistent with previous measurements for $La_{1-x}Sr_xCrO_3$[21] and $La_{1-x}Sr_xMnO_3$ for x=0.3. The difference between the X-ray absorption spectra measured with right (ρ+) and left (ρ-) circular-polarized light yields element-specific magnetic information for the Mn and Cr ions. The XMCD measurements were confirmed by switching the magnetic field direction. XMCD with a 0.5 Tesla in-plane field at 15 K are shown in Figure 3(a) and 3(b) at the Mn and Cr L-edges, respectively for a 3/3/3 sample. Dichroic signals were observed at the Mn and Cr edges indicating magnetic ordering on both the Mn and Cr sublattices. The XMCD hysteresis loop measured at the Mn L-edge in Figure 3(c) confirms that the LSMO layer is ferromagnetic. The dichroic signal at the Cr L-edge, while weak, exhibits an identical field dependence to the results at the Mn edge but in the opposite direction as shown in Figure 3(d). We conclude from the similarity of the hysteresis loops at the Cr and the Mn L-edges (Figure S2(a) in Supplementary Materials)[23] that the magnetism in the LSCO is an interfacial effect i.e. the magnetism in the LSCO layer is induced by the proximity to the ferromagnetic LSMO layer. The induced spin alignment anti-parallel to the applied magnetic field in the LSCO layers is confined to the LSMO/LSCO interface. The measured XMCD signal for Cr is five times smaller than Mn suggesting that the additional LSCO layers (away from the interface) in the 3/N/3 heterostructures are antiferromagnetic as expected for bulk LSCO.

The temperature dependence of the XMCD signal for a 3/6/3 heterostructure is shown in Figure 3(e) at the Mn L-edge. A reduction in the XMCD signal is observed as the temperature is increased from 80 K to 300 K. The magnitude of the XMCD intensities at the Cr and Mn L-edges are compared as a function of temperature in Figure 3(f) for the 3/6/3 sample. At both edges, a transition to a paramagnetic state is observed at ~ 200 K in agreement with the SQUID measurements in Figure 2(b) indicative of a coupling of the magnetic moments within the LSMO and LSCO layers.

*Role of Structural Coupling at the LSMO/LSCO Interface*

To determine the contribution of the atomic-scale structure to the enhanced magnetization, synchrotron surface diffraction experiments were performed at the 33ID beamline at the Advanced Photon Source to image the atomic-scale structures of the interface-engineered heterostructures. Crystal truncation rods (CTRs) and superstructure rods were measured to determine the atomic scale structures and octahedral rotations profiles. The crystalline quality of the 3/N/3 heterostructures is confirmed by the observance of clear Laue oscillations along the off-specular CTRs (Figure S3 in the Supplementary Materials)[23]. The CTRs were converted to real-space 3D electron density maps using the coherent Bragg rod analysis (COBRA)[20] phase retrieval technique from which the layer resolved atomic positions were extracted and refined using the GenX X-ray fitting algorithm[26]. From this analysis, the atomic positions were determined with sub-picometer resolution. The rotation of the oxygen octahedra in the LSMO and LSCO layers leads to a doubling of the perovskite unit cell and half order rods in reciprocal space. The amplitude of the octahedral rotations are determined from fits to the integer-order crystal truncation rods and the half-order rods measured (Figure S4 of the Supplementary Materials)[23] for the samples.

The converged structure for the 3/3/3 sample is shown in Figure 4(a). Out-of-phase ($a^-a^-c^-$ in the Glazer notation)[27] octahedral rotations are observed along the 3 crystallographic directions for the LSMO and LSCO layers. The in-plane B-O-B (B= Ti,Cr,Mn) bond angles for each layer is shown in Figure 4(b). The bond-angle decreases from 180º in the bulk STO layers to about 162º in the LSCO layers adjacent to the STO substrate. Within the LSMO layers, the average Mn-O-Mn bond angle is 166º in agreement with bulk LSMO. The reduced bond angles in the surface LSCO layers are attributed to a surface relaxation discussed below.

In addition to the octahedral rotations, cation-anion displacements are observed within the $BO_2$ planes. The layer-resolved displacements, $\Delta$, for the 3/3/3 sample are shown in Figure 4(c). Negative values of $\Delta$ indicate oxygen displacements towards the STO substrate relative to the cations z-positions. For bulk LSCO and LSMO, $\Delta=0$. However, $\Delta$ increases to an amplitude of 0.15 Å at top LSCO surface while the displacements in the LSMO layers are found to be suppressed.

To confirm that the suppressed distortions in the LSMO layers encapsulated with LSCO are related to the absence of polar discontinuities at the LSMO interfaces, we compare the layer-resolved lattice cation-anion displacements along the growth direction for a 4 uc LSMO film and a 3/4/3 heterostructure grown on (001) oriented SrTiO$_3$ are shown in Figure 5(a) and Figure 5(b), respectively. The measured crystal truncation rods and half-order Bragg peaks for the 3/4/3 sample are shown in Figure S5 and S6 respectively, of the Supplementary Materials. At the surface of the 4uc LSMO film and the LSCO cap layer of the 3/4/3 sample, the O anions are displaced toward the substrate relative to the cations. The ionic rumpling decays 3 uc below the film surface in agreement with previous results for a 10 uc LSMO film[14] and uncapped LaNiO$_3$ films.[20]. While polar distortions are observed in the 4 uc LSMO film, the distortions are suppressed in the encapsulated LSMO layers in the 3/3/3 and 3/4/3 samples.

The ionic displacements at the uncapped LSMO film surface are related to a surface electric field which arises due to the polar MnO$_2$ –terminated surface with a net -0.7$e$ charge. The decay length for the surface distortions are found to be on the order of 3 uc (~12 Å) which is consistent with the enhanced screening length observed for the rare-earth nickelates[20] and theoretical predictions for the manganites[15, 16]. For the uncapped LSMO film, the surface ionic rumpling leads to an elongation in the Mn-O bond-length which is correlated with reduced double exchange interactions and a suppression of ferromagnetic ordering in the uncapped LSMO film.[14, 16, 28]

On the other hand, the polar distortions are suppressed in the LSMO layers encapsulated with LSCO spacers leading to ferromagnetism in the LSCO/LSMO heterostructures. The measured layer resolved polar displacements for the 3/4/3 heterostructure shown in Figure 5(b) show that the polar distortions are confined to the top LSCO layer. The absence of polar distortions at the LSCO/STO substrate interface may be related to a reduction in the interface polar discontinuity driven by chemical intermixing observed in Figure 1(e). While the surface distortions may arise due to incomplete surface layers and atomic vacancies, the direction of the distortions point to the surface field as a significant driver for the observed distortions.[14, 20, 29]

The structural measurements and XMCD results confirm that the insertion of LSCO spacer layers leads to bulk-like Mn-O bonding and stoichiometric LSMO layers which are correlated with the stabilization of ferromagnetization in the encapsulated LSMO layers. The elimination of magnetic dead layers in the LSCO/LSMO/LSCO heterostructures is in contrast to the analogous nominally valence-matched $La_{0.6}Sr_{0.4}FeO_3$(LSFO)/$La_{0.6}Sr_{0.4}MnO_3$ interface where dead layers still exist.[30] At the LSFO/LSMO interface, an inherent charge transfer between Fe and Mn occurs which hole dopes the LSMO layers leading to antiferromagnetic interfacial LSMO layers. [30] This transfer, is reduced at the LSCO/LSMO interface, as evidenced by the ferromagnetic ground state of the 3/N/3 heterostructures.

The magnetization measured by SQUID normalized to the LSMO thickness in Figure 2(a) and the transition temperature of the 3/N/3 heterostructures are significantly less than the bulk values of 3.7 $\mu_B$/Mn and 370 K respectively. A reduced net SQUID magnetic moment is expected due to the net negative contribution of the magnetic moments in the LSCO layers measured by XMCD in addition to the effects of reduced dimensionality of the LSMO layers[31] and the effect of epitaxial strain.[32]

*Determination of Mn and Cr moments in [LSCO/LSMO] superlattices*

For the trilayer samples, the Ti-Cr intermixing at the STO/LSCO interface and the surface distortions observed in the surface LSCO layers are expected to affect the contribution of the LSCO layers to the total magnetization. Hence, to determine the layer-averaged magnetic moments of the LSMO and the LSCO layers, [LSCO (2)/LSMO (N)] superlattice samples were grown by molecular beam epitaxy on (001)-oriented STO substrates where the contributions of the interfacial and surface LSCO layers to the total magnetization is minimized. Representative STEM-EDS analysis of the [2/3]x8 sample is shown in Figure 6(a) indicating a well ordered structure with Cr-Mn intermixing on the order of a unit cell. Crystal truncation rod measurements along the specular and non-specular rods indicate coherently strained epilayers (Figure S7 of Supplementary Materials)[23].

The magnetization for the [LSCO (2)/LSMO (N)] superlattices measured by SQUID normalized to the LSMO thickness are shown as a function of applied magnetic field at 10 K in Figure 6(b) and temperature in Figure 6(c). The total SQUID magnetization per Mn, m*, decreases with the Mn thickness. By assuming that the total SQUID magnetization is a sum of the moments in the LSMO and LSCO sublattices, $m^* = m^{Cr}\left(\frac{M(LSCO)}{N(LSMO)}\right) + m^{Mn}$ where $m^{Cr}$ is the average moment per interfacial Cr ion in LSCO, $m^{Mn}$ is the average moment per Mn ion in LSMO, and M(LSCO) and N(LSMO) are the number of LSCO and LSMO layers, respectively. From the linear fit to the measured m* as a function of $\frac{M(LSCO)}{N(LSMO)}$ shown in Figure 6(d) for M=2 and N≤5, we determine the vertical intercept, $m^{Mn}$ =3.4 $\mu_B$/Mn which is close to the bulk value of 3.7 $\mu_B$/Mn and the slope, $m^{Cr}$ =-2.1 $\mu_B$/Cr.

A reduction in $m^{Mn}$ can be expected if confinement and the observed interfacial magnetic interactions lead to antiferromagnetic ordering and/or spin canting within the LSMO layers[19, 30, 33]. The model for the LSCO/LSMO heterostructure assumes that $m^{Mn}$ and $m^{Cr}$ are independent of the film thickness for M≤2 and N≤ 5. The linear dependence of m* on $\frac{M(LSCO)}{N(LSMO)}$ in Figure 6(d) is consistent with this assumption. Additionally, we find that the Mn XMCD signal for a 3/2/3 and 3/3/3 heterostructure (Figure S2(b)) are identical, in agreement with the assumption above.

To further validate the model described above, we compare two LSCO/LSMO heterostructures where the ratio of LSMO thickness to the LSCO thickness is 2:1. The magnetization as a function of temperature and magnetic field is measured for [LSCO (1)/LSMO (2)]x10 and [LSCO (2)/ LSMO (4)] x6 superlattice (see Figure S8). The transition temperatures and total magnetic moments are very close in these two configurations. Considering a moment of -2.1 $\mu_B$/Cr and a 3.4 $\mu_B$/Mn obtained from the linear fit in Figure 6(d) gives m*=2.245 $\mu_B$/Mn for [1/2] *10 and m*=2.17 $\mu_B$/Mn for [2/4] *6 superlattice in agreement with the measured magnetization in Figure S8 of the Supplementary materials.[23]

*First Principles Theory*

To understand the magnetic coupling of the LSCO (M)/LSMO (N) heterostructures, we perform DFT calculations using the Perdew–Burke–Ernzerhof exchange-correlation formalism[34] using Quantum Espresso[35] and ultrasoft pseudopotentials[36]; we estimate the magnetic moment of the various heterostructures by comparing the Lowdin projected magnetic moment for the bulk - for which the total available spin is known – with the moment in the superlattices. (For details on the calculations, the estimation of the magnetic moment, and the calculation's limitations, see the Materials and Method section.)

In order to focus on the interfacial magnetism, we perform calculations for a [2/2] superlattice using a collinear spin model imposing a FM state in LSMO as well as a FM state in LSCO, however with the two materials' spins aligned and anti-aligned. We find that the antiferromagnetic (AF) coupled state (i.e. LSMO spins anti-aligned with LSCO) is ~50 meV lower in energy compared to the FM coupled state (i.e. LSMO spins aligned with LSCO), in agreement with the experimentally determined antiferromagnetic coupling between the LSMO and LSCO layers. Next, we provide an estimate for the change in the Mn magnetic moment due to heterostructuring. In bulk FM LSMO, the spin available per Mn in a formula unit of LSMO is is predicted to be 3.7 $\mu_B$/Mn and for bulk LSCO the Cr available moment is estimated to be 2.7 $\mu_B$/Cr. For the [2/2] fully FM ordered superlattice we estimate a Mn moment of 3.61 $\mu_B$/Mn and 2.83 $\mu_B$/Cr in the [2/2] calculation, while for the calculation in which the two materials have their spins anti-aligned with each other but FM within the same material, we estimate 3.48 $\mu_B$/Mn and -2.81 $\mu_B$/Cr.

The theoretical Cr magnetic moment is overestimated compared to the experiment: this is likely because our calculations are done by treating the spins in the approximation that they are collinear (rather than non-collinear)[37]. For the experimental results discussed above, it is likely the spins on the Cr atoms suffer from frustration due to a competition between two magnetic interactions.

The first coupling is across the interface with the Mn where the interfacial Cr atoms are anti-aligned with the FM Mn atoms at the interface. The second effect is the coupling within LSCO itself which in the bulk is of AFM-G type[17].

**Discussion**
Based on the Goodenough-Kanamori rules, a FM superexchange interaction is expected between $Mn^{3+}(d^4, t_{2g}^3 eg^1)$ and $Cr^{3+}(d^3, t_{2g}^3 eg^0)$ due to $eg^0$-$eg^1$ interactions as has been observed in $LaMn_xCr_{1-x}O_3$[38]. Confinement and the tensile epitaxial strain imposed by the STO substrate will lead to a lifting of the degeneracy of the Mn eg orbitals where the half-filled $Mn^{3+}$ $eg(x^2-y^2)$ point towards the in-plane $Mn^{4+}$ ($d^3, t_{2g}^3 eg^0$) orbitals and the empty $eg(3z^2-r^2)$ orbitals point out-of-plane towards the LSCO layers. The in-plane $Mn^{3+}$-$Mn^{4+}$ exchange will be FM while the interfacial superexchange interaction between the empty $Mn^{4+}$ $eg(3z^2-r^2)$ orbital and the $Cr^{3+}$ will be AF as experimentally observed. This scenario is consistent with reports on bulk $La_{0.7}Ca_{0.3}Mn_{1-x}Cr_xO_3$ - where a significant decrease in the net magnetic moment is observed on increasing the Cr content.[39] Competition between these interactions leads to a net AFM coupling between the interfacial LSMO and LSCO layers as evidenced by the XMCD results and exchange bias demonstrated by hysteresis loops.[40, 41] The competing interactions is consistent with a decrease in the Curie temperature as the FM LSMO thickness is reduced.

We note that reduced moments in the thinner LSMO layers may be attributed to charge transfer between the Cr and Mn ions at the LSCO/LSMO interface. To investigate the Cr charge states, we measured the chemical shifts of X-ray absorption near-edge spectra at the Cr K-edge of the [LSCO(2)/LSMO(5)] heterostructure relative to a Cr metal standard (see Figure S9 of the Supplementary Materials). The Cr oxidation state is determined to be +3.3 +/- 0.1 which is expected for a stoichiometric $La_{0.7}Sr_{0.3}CrO_3$.[42] However, within the experimental error, it is likely that the transfer of holes from Cr to Mn will lead to a reduction of the Mn moment and an increase in the $Cr^{3+}$ content at the interface further favoring the AF Mn-Cr exchange. The Cr ELNES (Figure S10) averaged for each layer in the superlattice remains unchanged through the thickness of the film, supporting XAS oxidation state observations. [43] Additionally, we do not observe chemical shifts at the Mn K-edge for the [LSCO (M)/LSMO (N)] superlattices compared to a bulk-like 30 uc LSMO film which confirms that the growth conditions and heterostructuring do not lead to significant changes in the stoichiometry and charge states of the LSMO layers. This observation is confirmed by the EELS white line ratio for the [2/3]x8 samples in supplementary Figure S10[23] which shows little variation in the white line ratio across the film indicating a near constant Mn valence across the film. [43] [10, 30]

In conclusion, we have demonstrated the confinement of magnetism in atomically thin LSMO layers by inserting spacer layers which effectively compensate the interfacial and surface polar discontinuities on (001)-oriented STO substrates. By choosing $LaSrCrO_3$ as a spacer layer, we also reduce the interfacial oxygen octahedral rotation mismatch[18] leading to bulk-like Mn-O bonding which favors long range ferromagnetic ordering in the LSMO layers. This approach effectively removes structural distortions related to the magnetically dead interfacial layers leading to enhanced ferromagnetism in 2 uc thick LSMO films encapsulated with LSCO spacers. XMCD measurements indicate antiferromagnetic coupling between the ferromagnetic LSMO films and the interfacial Cr layers which also contributes to the stabilization of 2D magnetism in the LSMO films. The interfacial antiferromagnetic coupling leads to a magnetic exchange bias as evidenced by SQUID measurements. The ability to engineer the electrostatic and structural boundary conditions at oxide interfaces has important implications for engineering exotic phenomena in two-dimensional oxide systems and the design of nanoscale devices. The observed relation between the interfacial polar distortions on the magnetic properties of the LSMO layers has important implications in the design of multiferroic materials where the electric field effect may be used to modulate the magnetization of the ultrathin oxide layers.

**Materials and Methods**

*Sample Preparation*

The LSCO layers and the LSMO films were grown at 800°C in 3x10$^{-6}$ Torr atomic oxygen from an oxygen plasma source on TiO2-terminated SrTiO$_3$ (Crystec) substrates by molecular beam epitaxy. The films were grown by co-deposition from effusion cells at a growth rate of approximately 1 unit cell per minute. After growth, the samples were slowly cooled down at 5 °C/min in 5x10$^{-6}$ Torr oxygen in the growth chamber to ensure complete oxidation. The film thickness and surface crystallinity were monitored *in-situ* by reflection high energy electron diffraction (RHEED).

*Electron Microscopy*

Electron microscopy samples were prepared by conventional cross-section mechanical polishing and argon ion milling. Aberration-corrected high angle annular dark field-scanning transmission electron microscopy (HAADF-STEM) imaging and energy-dispersive x-ray spectroscopy (EDS) were performed using a FEI Titan G2 60-300 kV operated at 200 kV with a convergence semi-angle of 19.6 mrad. The electron beam was monochromated to achieve 0.4 eV energy resolution and a dispersion of 0.05 eV/channel was used. Electron energy loss spectroscopy (EELS) was performed to determine the variation of Cr and Mn valence by tracking three spectral features across the LSCO/LSMO heterostructure: the energy-loss near edge fine structure (ELNES), the energy shift of the L-edges, and the $L_3/L_2$ integrated white line ratio (WLR)[22, 43]. EELS data were collected by acquiring 2-D spectrum images of the film at a sampling rate of approximately 1Å/pixel.

*SQUID Magnetization Measurements*

The magnetic properties of the LSMO thin films and M/N/M heterostructures and [M/N] superlattices were characterized using a Quantum Design SQUID system. The temperature dependent magnetization curves were measured on warming in an applied in-plane field of 1 KOe.

*Synchrotron Diffraction Measurements*

Synchrotron surface diffraction experiments were performed at the 33ID beamline at the Advanced Photon Source to image the atomic-scale structures of the interface-engineered heterostructures using an incident X-ray energy of 15.5 KeV. The samples were mounted in a Be-dome chamber and pumped down to less than $10^{-4}$ Torr. Crystal truncation rods (CTRs) and superstructure rods were measured on each sample and analyzed using the coherent Bragg rod analysis (COBRA) phase-retrieval technique. The atomic positions and occupations obtained from the COBRA-determined electron density maps were refined using the GenX genetic fitting algorithm. For each fitting parameter used in the structural refinement (atomic position, occupation, Debye Waller factor), the error bars were obtained by determining the change in the converged value which leads to a 5% increase in the optimal crystallographic R1 factor.

*Soft X-ray XAS and XMCD Measurements*

Soft X-ray magnetic circular dichroism (XMCD) measurements at the Mn and Cr $L_{2,3}$ absorption edges at were performed at the 4.0.2 beamline at the Advanced Light Source using total electron yield. A magnetic field of 0.5 Tesla was applied parallel to the sample surface.

*Density Functional Theory Methodology*

The DFT calculations have been performed using the Quantum Espresso code[35] using the PBE exchange correlation functional[34] and ultrasoft pseudopotentials[36] using a k-mesh of 7x7x5 for the bulk 20 atom unit cell and 7x7x3 for the 40 atom superlattices. For a c(2x2)x2 20 atom unit cell for LSMO we obtain an in-plane lattice constant of 3.90 Å and simulate the strain to the STO lattice constant by imposing an in-plane lattice constant of 3.905 Å on the superlattices.

For bulk LSMO we correctly obtain a ferromagnetic ground state. However, for the LSCO we obtain AFM-C instead of AFM-G as the ground state. For a 20 atom unit cell we find that the AFM-G state is 24 meV higher than AFM-C; while the AFM-A state is 86meV and FM state is 213 meV higher than AFM-C. It is likely that the issues in LSCO calculations are due to the fact that we are using a collinear spin calculation, as applying a U did not stabilize the AFM-G ground state and previous studies have used a non-collinear spin calculation to correctly predict the AFM-G ground state. Finally, we need a correct way to estimate the local magnetic moment of the atoms, as we notice that in bulk STO-strained LSMO, the magnetic moment calculated on the Mn site using Lowdin orbital projections is 3.35 $\mu_B$/Mn, a lower number than the experimentally expected one, and lower than what would obtain by dividing the total magnetization of the unit cell by the number of Mn sites, a number which is exactly 3.7 $\mu_B$/Mn. This is the due to the fact that the Lowdin d orbitals of Mn are not sufficient to obtain a complete representation of the valence bands of LSMO and the resulting magnetic moment. We also find that a 5 atom formula unit of LSCO has 2.7 $\mu_B$/Cr in a FM calculation, with the projected moment on the Cr atoms of 2.43 $\mu_B$/Cr.

To estimate the magnetic moment of the LSMO and LSCO in the [LSCO(2)/LSMO(2)] heterostructure, we consider the projected magnetic moment onto the d-shells. In the FM-coupled [LSCO(2)/LSMO(2)] heterostructure the projected moment of the Mn atom is 3.25 $\mu_B$/Mn and that of the Cr is 2.55 $\mu_B$/Cr, a small reduction in the magnetic moment of the Mn and a small increase in that of the Cr compared to the FM bulk calculations for the two materials. Assuming the change in magnetic moment obtained using Lowdin projectors is directly proportional to the change per unit cell, we estimate the moments per 5 atom cell to be 3.59 $\mu_B$/Mn and 2.83 $\mu_B$/Cr. Using this prediction, the prediction for the total magnetization for the heterostructure is 25.68 $\mu_B$, which compares well with the computed total magnetization of 25.60 $\mu_B$. Similarly, for the AFM computed structure, we obtain a projected moment of 3.16 $\mu_B$/Mn and 2.53 $\mu_B$/Cr which would translate into a moment of 3.49 $\mu_B$/Mn – similar to the experiment, and 2.81 $\mu_B$/Cr, higher than measured in the experiment. The error here is larger than in previous calculations: the total moment per calculation is 2.72 $\mu_B$/cell, while the calculated total moment is 3.11 $\mu_B$/cell, with an estimated error of 0.05 $\mu_B$/formula unit in the estimation.

**Acknowledgments**

**General**: A.B.G. acknowledges discussions with A.J. Millis and support from the Flatiron Institute's Scientific Computing Core.

**Funding:** D.P.K. and S.K. acknowledge financial support by the US National Science Foundation under Grant No. NSF DMR1751455. A.N.P and J.M.L acknowledge support from the National Science Foundation under Grant No. DMR-1608656.This research used resources of the Advanced Light Source, which is a DOE Office of Science User Facility under contract no. DE-AC02-05CH11231. Use of the Advanced Photon Source was supported by the U.S. Department of Energy, Office of Science, Office of Basic Energy Sciences, under Contract No. DE-AC02-06CH11357. The Flatiron Institute is a division of the Simons Foundation. This work was performed in part at the Analytical Instrumentation Facility (AIF) at North Carolina State University, which is supported by the State of North Carolina and the National Science Foundation (award number ECCS-1542015). The AIF is a member of the North Carolina Research Triangle Nanotechnology Network (RTNN), a site in the National Nanotechnology Coordinated Infrastructure (NNCI).

**Author contributions:** D.P.K. conceived and supervised the project. D.P.K. and SK. performed the synchrotron diffraction measurements and analyzed the data. S.K. fabricated the samples and performed the magnetization measurements. A.B.G performed the theoretical calculations. E.A. and D.P.K. performed the synchrotron spectroscopy measurements and analyzed the data. A.N.P. and J.M.L. performed the electron microscopy measurements.

**Competing interests:** There are no competing financial interests of the authors.
**Data and materials availability:** Additional data related to this paper may be requested from the authors.

**Figures and Tables**

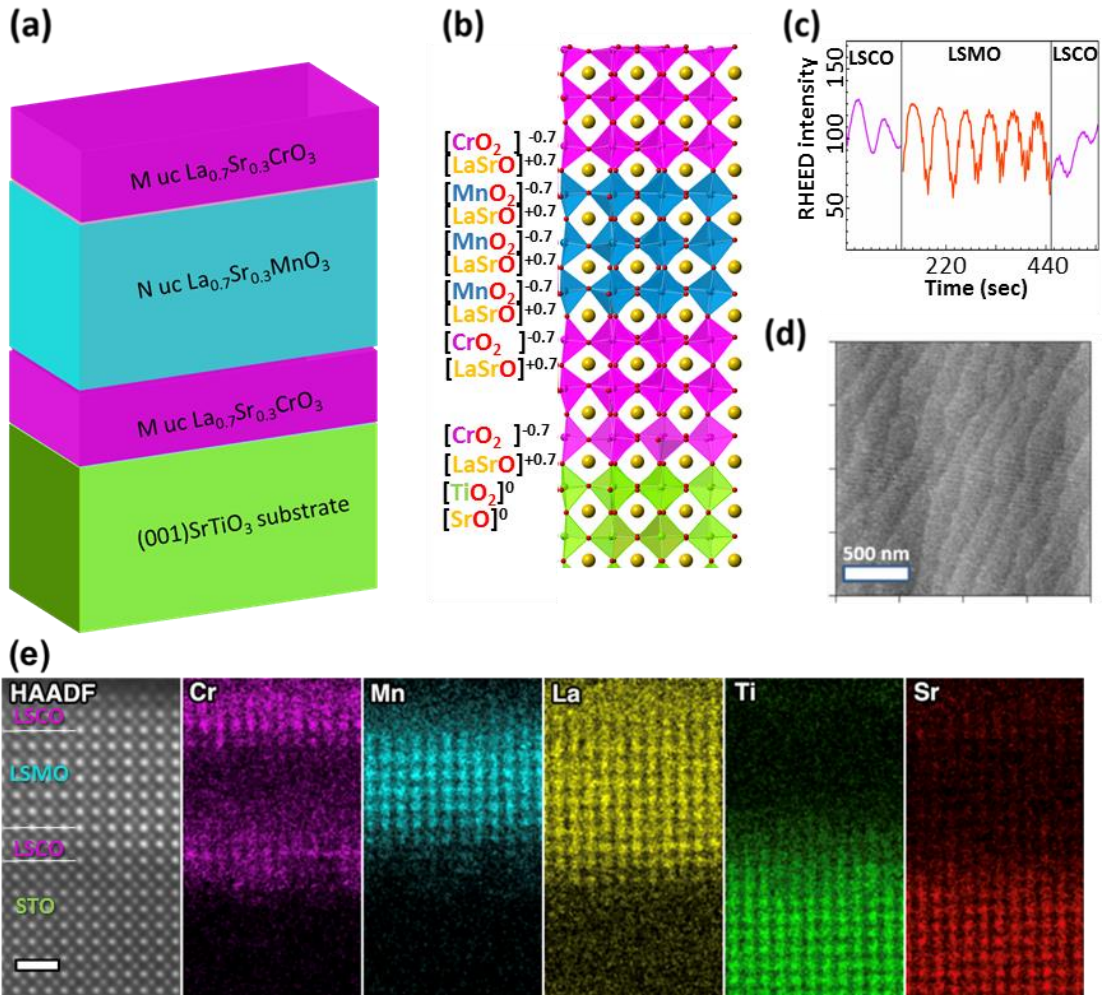

**Figure 1: Tuning the valence-mismatch at manganite interfaces**. (**a**) Schematic of $La_{0.7}Sr_{0.3}CrO_3$ (M)/ $La_{0.7}Sr_{0.3}MnO_3$ (N)/$La_{0.7}Sr_{0.3}CrO_3$ (M) (M/N/M) heterostructures grown by molecular beam epitaxy on (001)-oriented $SrTiO_3$ substrates. (**b**) Schematic of a 3/3/3 structure showing the net charges in the AO (A=La,Sr) and $BO_2$ (B=Ti,Mn,Cr) layers. (**c**) RHEED oscillations recorded in-situ during the growth of a 2/6/2 heterostructure. (**d**)The corresponding atomic force microscope image of the surface topography of the 3/3/3 heterostructure exhibiting unit cell high terrace steps. (**e**) Representative HAADF-STEM image and EDS map of nominally 2/6/2 LSCO/LSMO. The scale bar represents 1 nm.

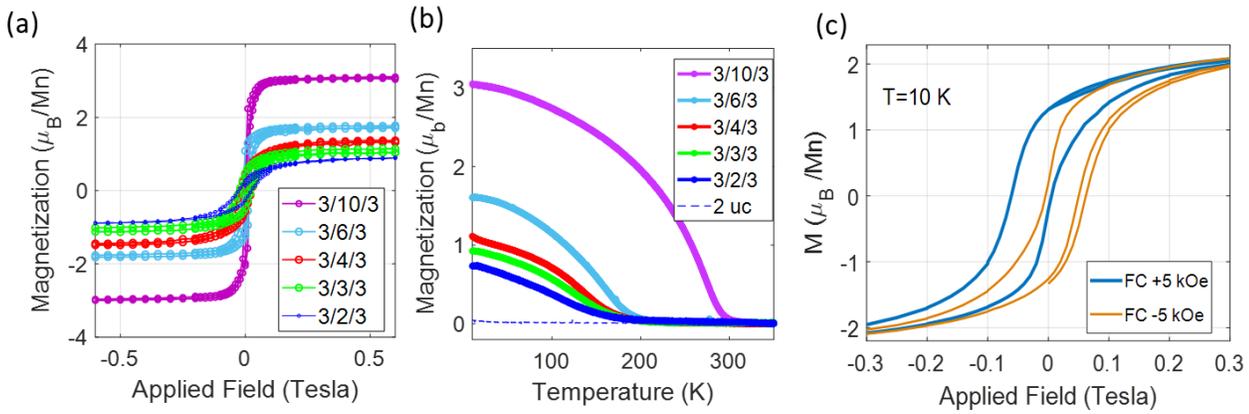

**Figure 2: Effect of interface engineering on ferromagnetic ordering in LSCO (3)/LSMO (N)/LSCO (3) heterostructures**. (**a**) Total magnetic moment per Mn as a function of applied magnetic field for 3/N/3 LSCO structures measured at 10 K. The diamagnetic contribution of the SrTiO$_3$ substrate has been subtracted. (**b**) Comparison of temperature dependent magnetization for a 2 uc LSMO film and 3/N/3 heterostructures grown on (001) oriented SrTiO$_3$ where N=2,3,4,6,10 unit cells. The curves are measured on warming in an applied in-plane field of 1 KOe. (**c**) Magnetic hysteresis loops of a [LSCO (2)/LSMO (4)]x6 superlattice measured at 10 K by SQUID. The sample was cooled down from the room temperature in the presence of ±0.5 T magnetic field.

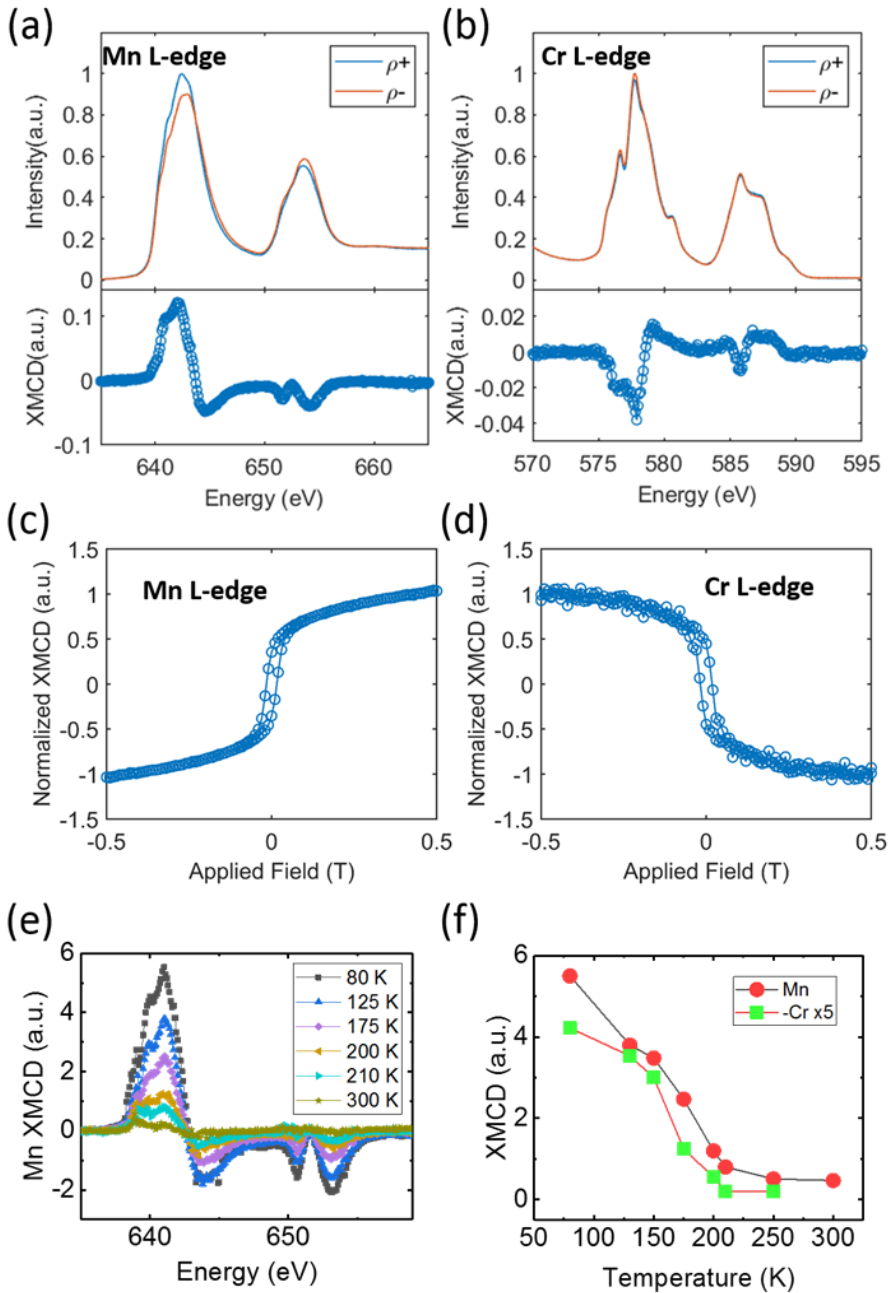

**Figure 3: Magnetic properties of LSCO (M)/LSMO (N)/LSCO (M) (M/N/M) heterostructures probed by X-ray magnetic circular dichroism (XMCD) in total electron yield mode.** XMCD results for a 3/3/3 heterostructure measured at the (**a**) Mn L-edge and the (**b**) Cr L-edge at 15 K. Normalized applied magnetic field dependence of the XMCD signal measured at (**c**) the Mn L-edge and (**d**) the Cr L-edge for the 3/3/3 sample. (**e**) Temperature-dependent XMCD results for a 3/6/3 heterostructure at the Mn L-edge. (**f**) Comparison of the magnitude of temperature-dependent XMCD signal at the Mn and Cr L edges in an 0.5 Tesla magnetic field applied in the plane of the sample for a 3/6/3 sample. The Cr data is multiplied by a factor of 5 for clarity.

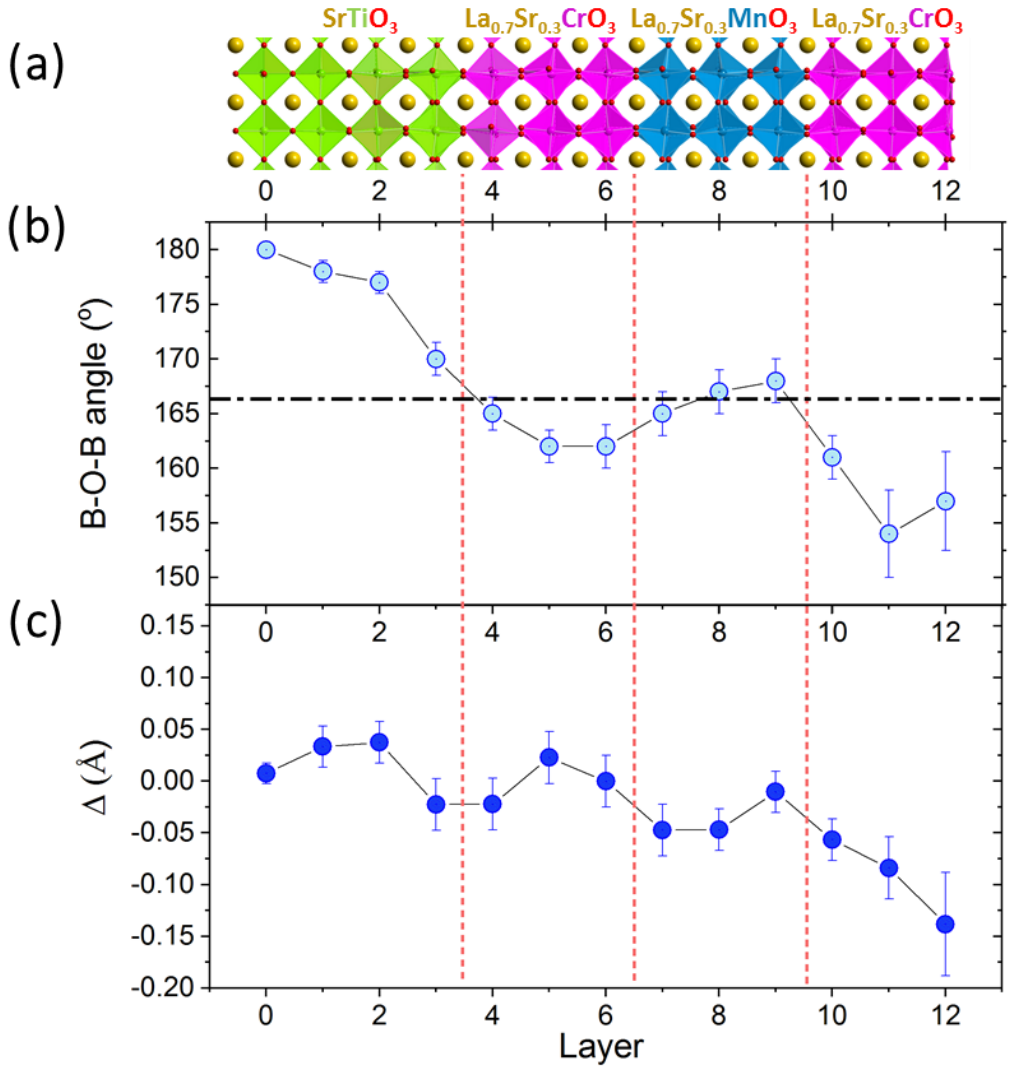

**Figure 4: Atomic-scale structure of a LSCO (3)/LSMO (3)/LSCO (3) heterostructure. (a)** Layer-resolved Oxygen octahedral rotations of a 3/3/3 heterostructure on SrTiO$_3$ determined from fits to crystal truncation rods and half-order superstructure rods. (**b**) Layer-resolved in-plane B-O-B (B=Ti layers 0-3, Cr layers 4-6, Mn layers 7-9, Cr layers 10-12) The dashed line indicates the bulk Mn-O-Mn bond angle for LSMO. (**c**) Layer-resolved B-O displacements, Δ, along the growth direction for the 3/3/3 heterostructure. Δ represents the displacements in Angstroms, of the oxygen ions relative to the cation (Ti,Mn,Cr) z positions in each layer. Positive Δ are toward the film surface.

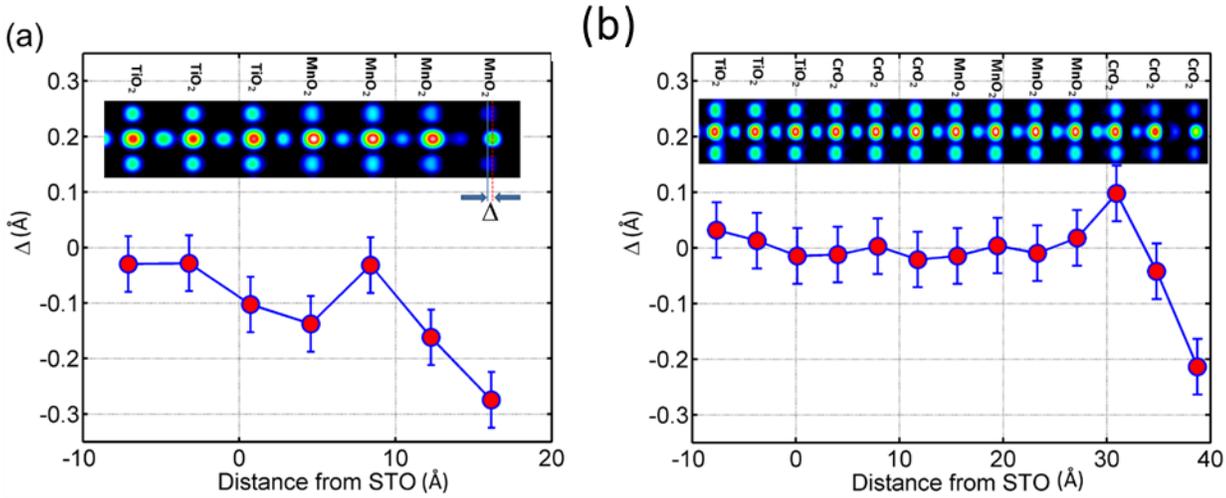

**Figure 5: Layer-resolved polar distortions for LSMO films and LSCO (3)/LSMO (4)/LSCO (3) heterostructures determined from synchrotron X-ray diffraction crystal truncation rod measurements.** Layer-resolved ionic displacements for **(a)** a 4 unit cell $La_{0.7}Sr_{0.3}MnO_3$ film and **(b)** a 3/4/3 sample obtained from synchrotron X-ray crystal truncation rod measurements. The insets show a cut through the measured electron density profiles showing the $BO_2$ planes and the apical O ions. $\Delta$ represents the displacements in Angstroms, of the oxygen ions relative to the cation (Ti,Mn,Cr) z positions. Positive $\Delta$ are toward the film surface.

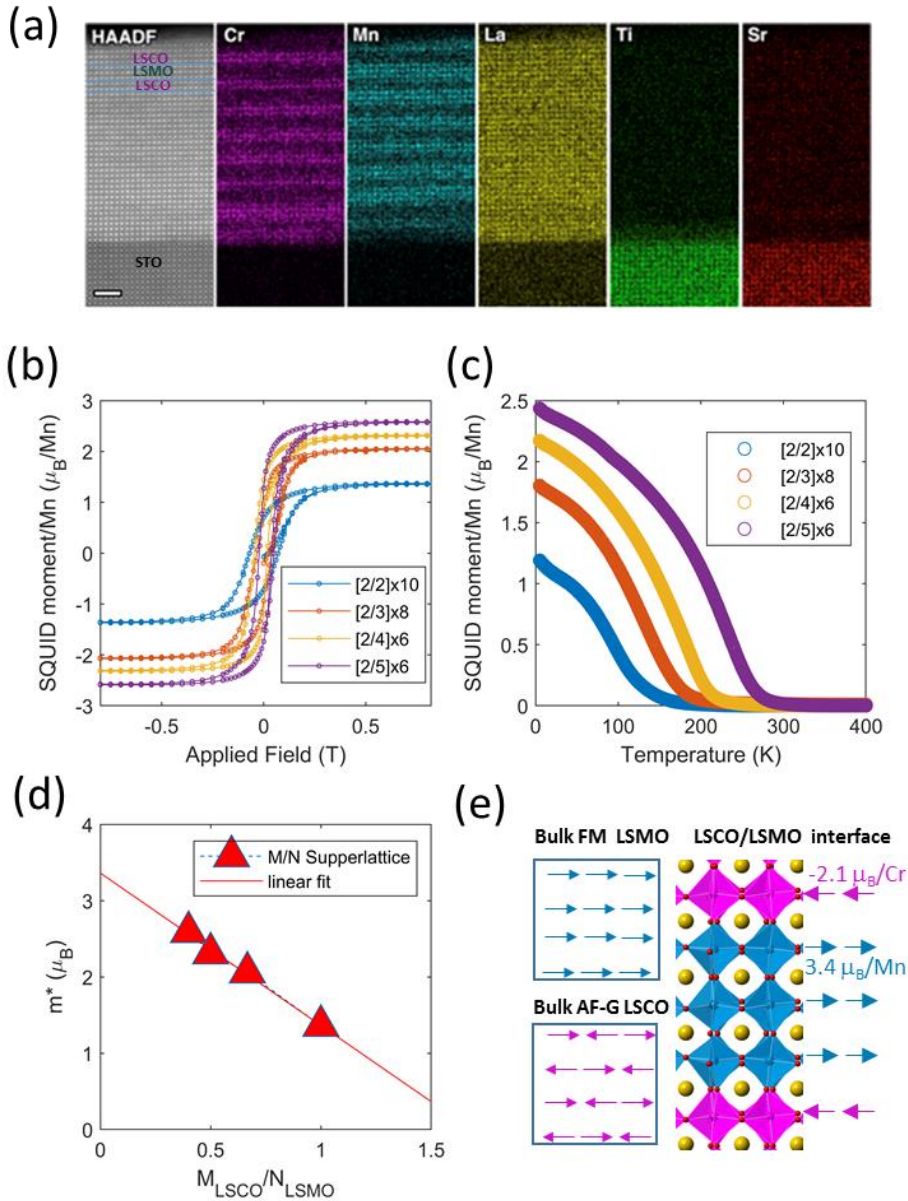

**Figure 6: Magnetic properties of [LSCO (M)/ LSMO (N)] superlattice samples.** (**a**) HAADF-STEM image and EDS map of [2/3]x8 sample. The scale bar represents 2 nm. (**b**) Field dependence at 10 K of the magnetic moment measured by SQUID normalized to the total LSMO thickness. (**c**) Temperature dependence of the SQUID magnetization normalized to the total LSMO thickness (**d**) SQUID magnetization at 10 K normalized to the total LSMO thickness, m*, as a function of the ratio of the LSCO and LSMO thickness. The magnetic moment per Mn is determined from the intercept of the linear fit with the vertical axis to be 3.4 +/0.1 $\mu_B$/Mn. The magnetic moment per Cr is determined from the slope of the linear fit to be -2.1+/0.1 $\mu_B$/Cr. (**e**) Model of the magnetization for bulk LSMO and LSCO and along the growth direction in the LSCO/LSMO heterostructures below the Curie temperature.

# Supplementary Information for "Confinement of magnetism in atomically-thin La0.7Sr0.3CrO3/La0.7Sr0.3MnO3 heterostructures"

Sanaz Koohfar[1], Alexandru Bogdan Georgescu[2], Aubrey Penn[3], James M. LeBeau[3], Elke Arenholz[4], Divine Philip Kumah[1]

**Affiliations**
[1]Department of Physics, North Carolina State University, Raleigh, NC, 27695, USA.

[2]Center for Computational Quantum Physics, Flatiron Institute, 162 5th Avenue, New York, NY 10010, USA.

[3]Department of Materials Science and Engineering, North Carolina State University, Raleigh, NC, 27695, USA.

[4]Advanced Light Source, Berkeley, CA, 94720, USA

**Supplementary Discussion**

*Magnetic Measurements*

An important issue in measuring the magnetic signals of ultra-thin films is determining the contribution of the underlying substrate. Bulk $SrTiO_3$ is expected to be diamagnetic however, the presence of magnetic impurities on the ppm level and high temperature annealing conditions can lead to finite ferromagnetism and paramagnetic effects. Figure S1 show the uncorrected magnetization versus temperature for a bare STO, a 2 uc LSMO film, a 6 uc LSCO film and a LSCO(3)/LSMO(2)/LSCO(3) (3/2/3) heterostructure grown on STO. It is evident from the raw data that a ferromagnetic signal sits atop a paramagnetic background only for the 3/2/3 heterostructure. The ferromagnetic ordering in the 3/2/3 heterostructure is confirmed by the magnetization versus field measurements as shown in Figure 2 of the main text.

**Structural refinement**

The atomic coordinates, layer composition and Debye-Waller factors extracted from the COBRA-derived electron densities are refined using the GenX fitting program. The layer resolved oxygen-octahedral tilts and rotations are determined by fitting the measured integer-order crystal truncation rods (Figure S3) and the half-order rods (Figure S4) for the 3/3/3 sample. The corresponding rods are shown in Figure S5 and Figure S6 for the 3/4/3 sample. For the fit, we assume equal fractions of 4 90° rotation domains.

**Effect of growth conditions on magnetization**

Two growth procedures were compared for the synthesis of the LSCO (M)/LSMO (N) heterostructures by MBE. In the first approach, the LSCO (M)/LSMO (N) heterostructures were LSCO layers were grown in $10^{-8}$ Torr molecular oxygen and the LSMO films were grown in $3 \times 10^{-6}$ Torr atomic oxygen from a plasma source. The samples were cooled down in $1 \times 10^{-6}$ Torr oxygen in the growth chamber at a rate of 25°C/minute. Film growth was followed by an ex-situ post anneal in high purity $O_2$ at 700 °C for 12 hours to minimize the formation of Oxygen vacancies. The second approach involved the growth of both the LSCO and LSMO layers at in $3 \times 10^{-6}$ Torr atomic oxygen from a plasma source. The samples were cooled down at 5 °C/min in $5 \times 10^{-6}$ Torr oxygen in the growth chamber to ensure complete oxidation. We find that while both growth methods result in ferromagnetic ground states, the second approach leads to higher paramagnetic-ferromagnetic transition temperatures and higher net moments. The differences may be attributed to differences in oxygen stoichiometry of the LSMO layers. For the first method, the low oxygen pressure used for the LSCO growth may result in a slight reduction of the LSMO layers which is avoided by growing both layers in a high oxygen pressure.

**Electron energy-loss spectroscopy (EELS)**

EELS was performed to map the variation of Cr and Mn valence by tracking three spectral features across the LSCO/LSMO heterostructure: the energy-loss near edge fine structure (ELNES), the energy shift of the L-edges, and the $L_3/L_2$ integrated white line ratio (WLR)[22, 43]. EELS data were collected by acquiring 2-D spectrum images of the film at a sampling rate of approximately 1Å/pixel (additional experimental parameters are given in the main text). Figure S10(a) shows an EELS elemental color map and a HAADF-RevSTEM [44] image overlaid, revealing the alternating

LSCO and LSMO layers. Rows of the spectrum images were averaged to achieve the Cr and Mn L-edge intensity images shown in Figures S10 (b,c), with the distance from the interface corresponding directly to Figure S10 (a). There is negligible change in the energy shift of both the Cr and Mn edges. The intensities corresponding to peak fine structure show oscillations in L-edge intensity with B-site composition change, that combined with the unchanged energy shift, indicates a constant B-site valence. L-edge intensities change gradually between layers, supporting the EDS observations that some Cr-Mn intermixing occurs in the heterostructure. Additionally, near the interface, the overall intensity of Cr edge decreases, consistent with the Ti B-site substitution shown in EDS results. The WLR of Mn and Cr (shown in Figure S10 (d) also as a function of film depth) was determined following the process outlined by Tan, et al.[22] with a custom Matlab script including Fourier-ratio deconvolution to remove plural scattering effects, a two-step background correction, and integration of the peaks with a 4 eV integration window per peak. The WLR is related to valence by an exponential function, but the values are roughly inversely proportional. The Cr and Mn WLR remain mostly unchanged throughout the heterostructure which, combined with shifting and ELNES results, confirms a constant B-site valence.

**Supplementary Figures**

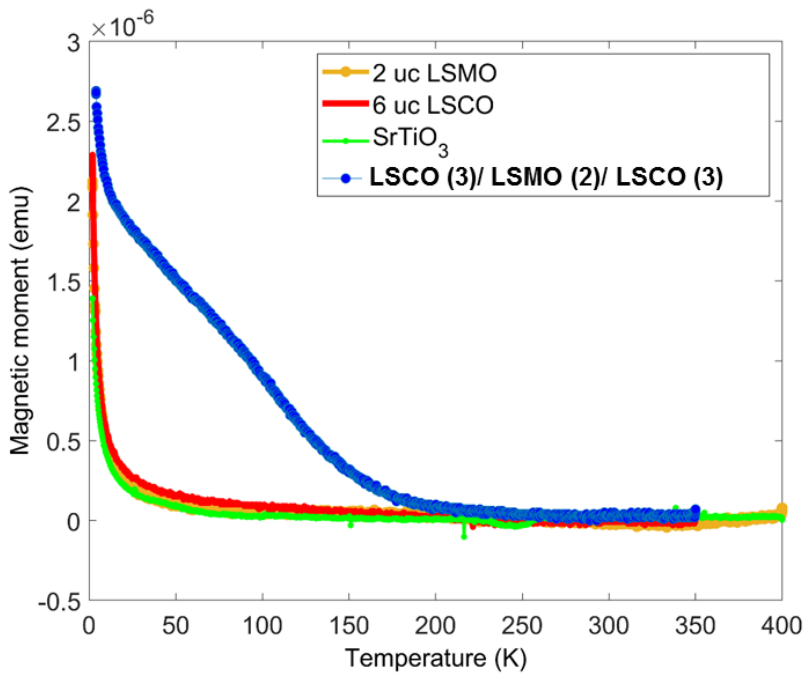

**Figure S1: Comparison of magnetic moments measured by SQUID for a 2 unit cell La$_{0.7}$Sr$_{0.3}$MnO$_3$ film, a 6 uc La$_{0.7}$Sr$_{0.3}$CrO$_3$ film, an SrTiO$_3$ substrate and a La$_{0.7}$Sr$_{0.3}$CrO$_3$(3)/La$_{0.7}$Sr$_{0.3}$MnO$_3$ (2)/La$_{0.7}$Sr$_{0.3}$CrO$_3$(3) heterostructure grown on (001) oriented SrTiO$_3$ by molecular beam epitaxy.** Measurements were taken while heating up the samples an applied field of 500 Oe after field cooling in 1000 Oe. For comparison, the minimum magnetization for each curve has been set to 0 emu by subtracting a constant background. The substrate contribution has not been subtracted. A paramagnetic-ferromagnetic transition is observed for the La$_{0.7}$Sr$_{0.3}$CrO$_3$(3)/La$_{0.7}$Sr$_{0.3}$MnO$_3$ (2)/La$_{0.7}$Sr$_{0.3}$CrO$_3$(3) heterostructure at ~150 K.

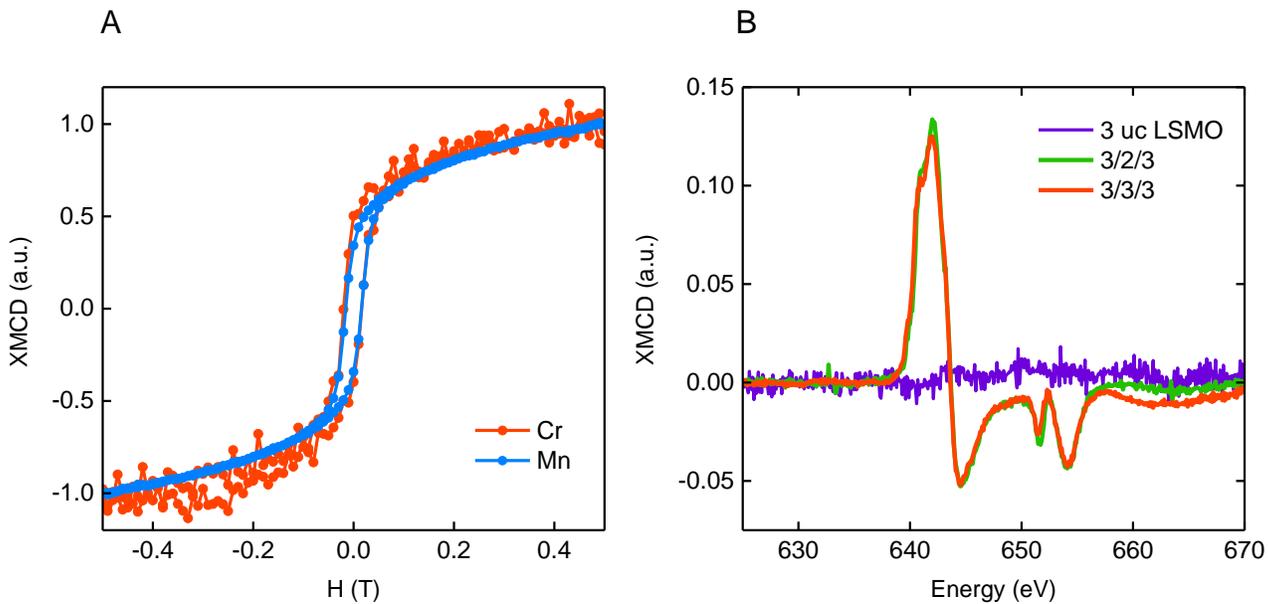

**Figure S2: (a) Comparison of field-dependent normalized ferromagnetic XMCD signal measured at the Mn L edge and the negative signal measured at the Cr L edge for a LSCO (3) /LSMO (3) /LSCO (3) heterostructure.** The field is applied in the plane of the sample. The overlap of the two curves indicates antiferromagnetic coupling of the magnetization in the LSMO layer with magnetism in the LSCO layer. **(b) XMCD signal measured at Mn L edge at 15 K edge for a 3 uc LSCO/2 uc LSMO/3 uc LSCO and a 3 uc LSCO/3 uc LSMO/3 uc LSCO. XMCD signal for a 3 uc LSMO film on STO is shown for comparison.**

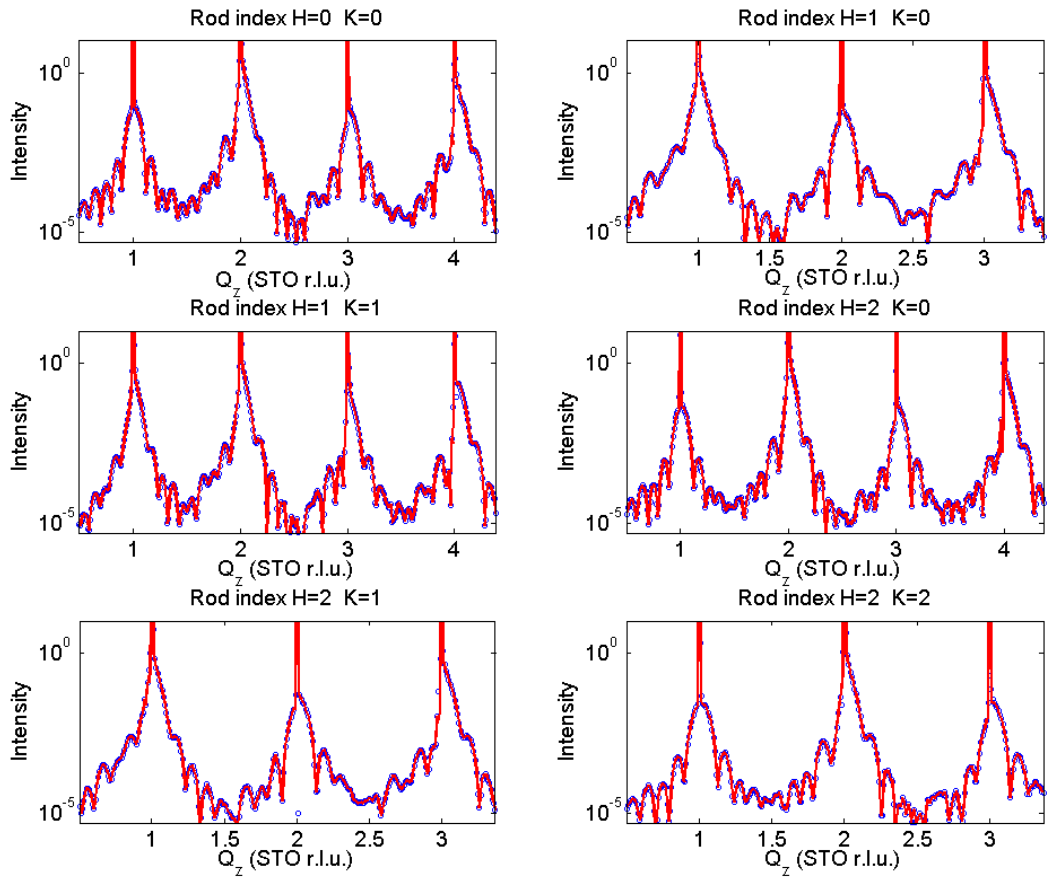

**Figure S3:** Measured crystal truncation rods (blue circles) and best fit structure (red lines) for a LSCO (3)/LSMO (3)/ LSCO (3) heterostructure grown on $SrTiO_3$. 1 STO reciprocal lattice unit (r.l.u)=1/3.905 $Å^{-1}$.

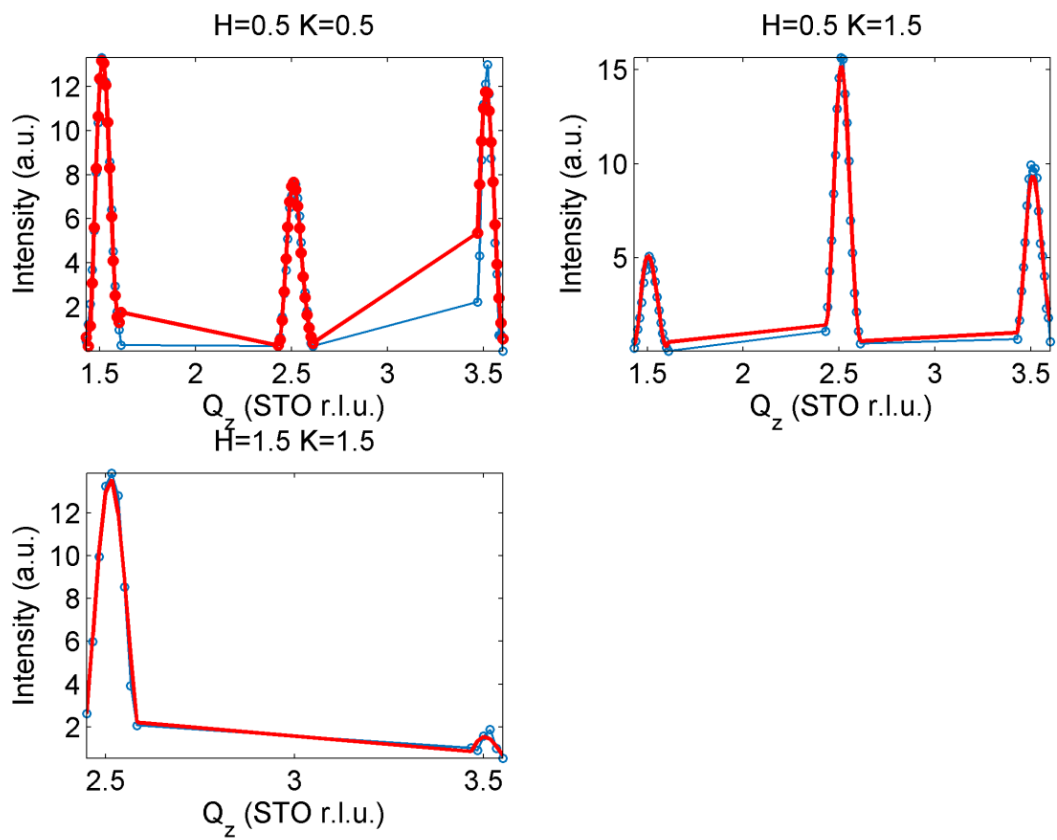

**Figure S4: Measured half order rods (blue circles) and best fit structure (red lines) for a LSCO (3)/LSMO (3)/ LSCO (3) heterostructure grown on SrTiO$_3$.**

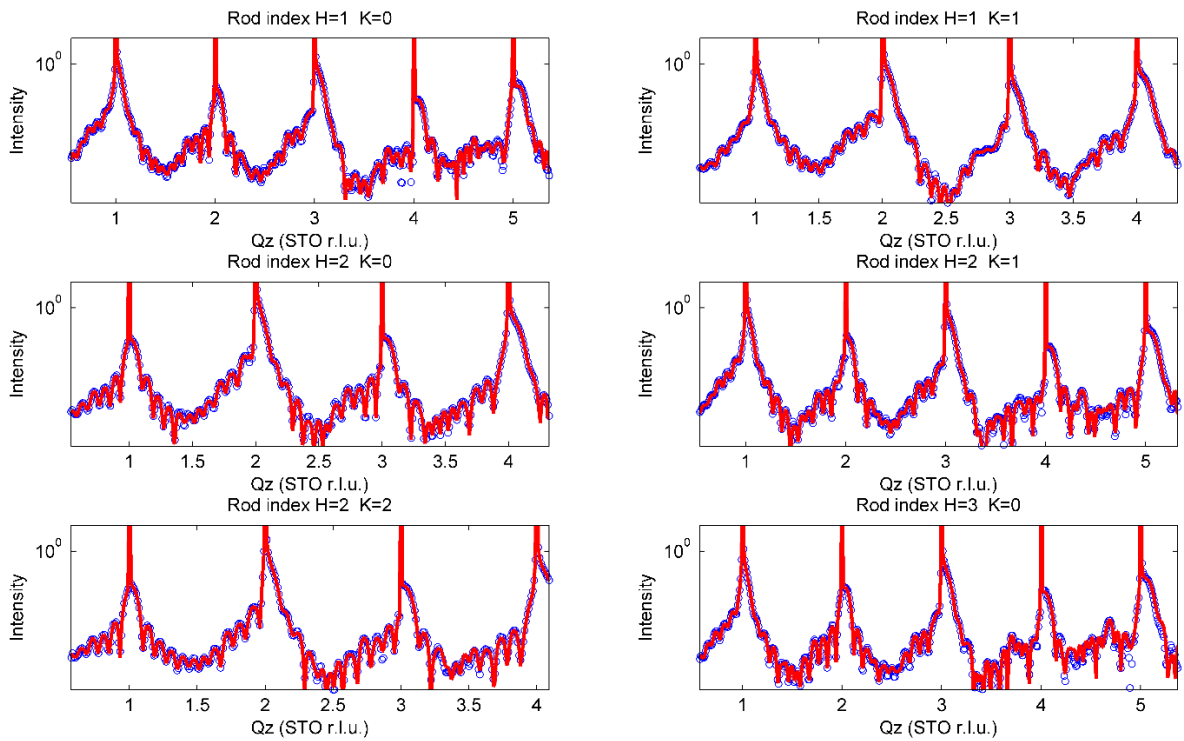

**Figure S5: Measured crystal truncation rods (blue circles) and best fit structure (red lines) for a LSCO (3)/LSMO (4)/ LSCO (3) heterostructure grown on SrTiO$_3$. 1 STO reciprocal lattice unit (r.l.u)=1/3.905 A$^{-1}$.**

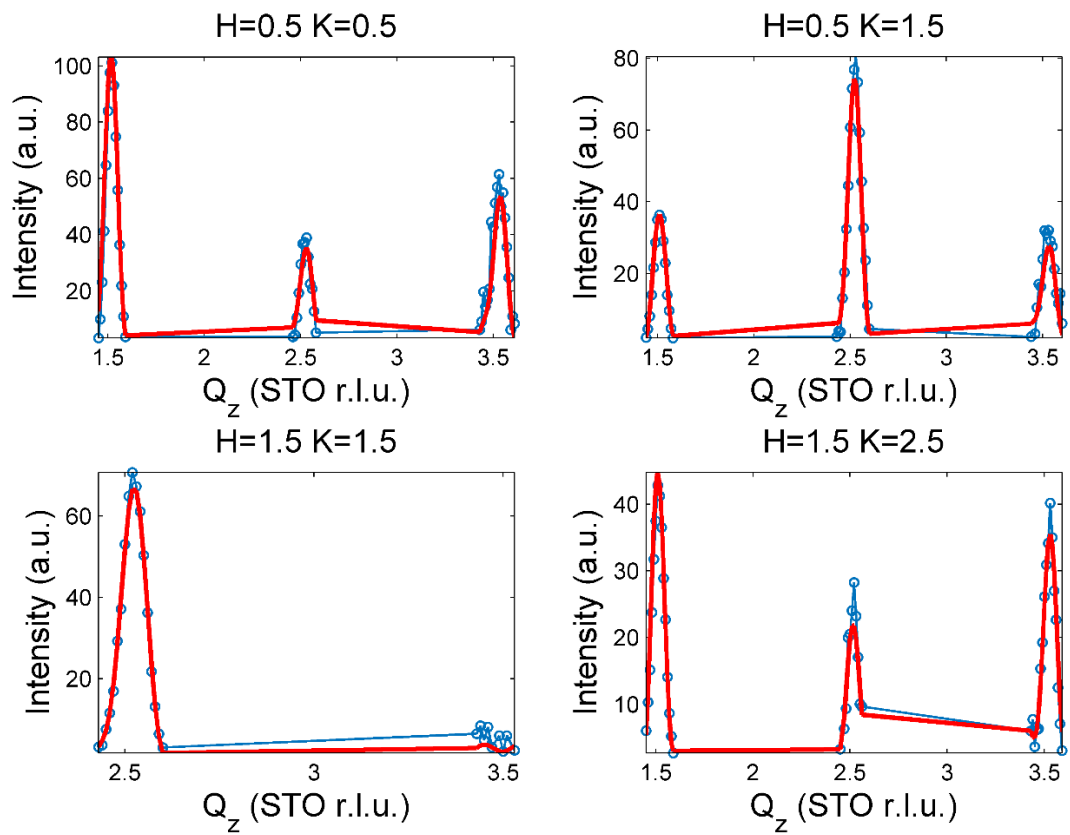

**Figure S6: Measured half order rods (blue circles) and best fit structure (red lines) for a LSCO (3)/LSMO (4)/ LSCO (3) heterostructure grown on SrTiO$_3$.**

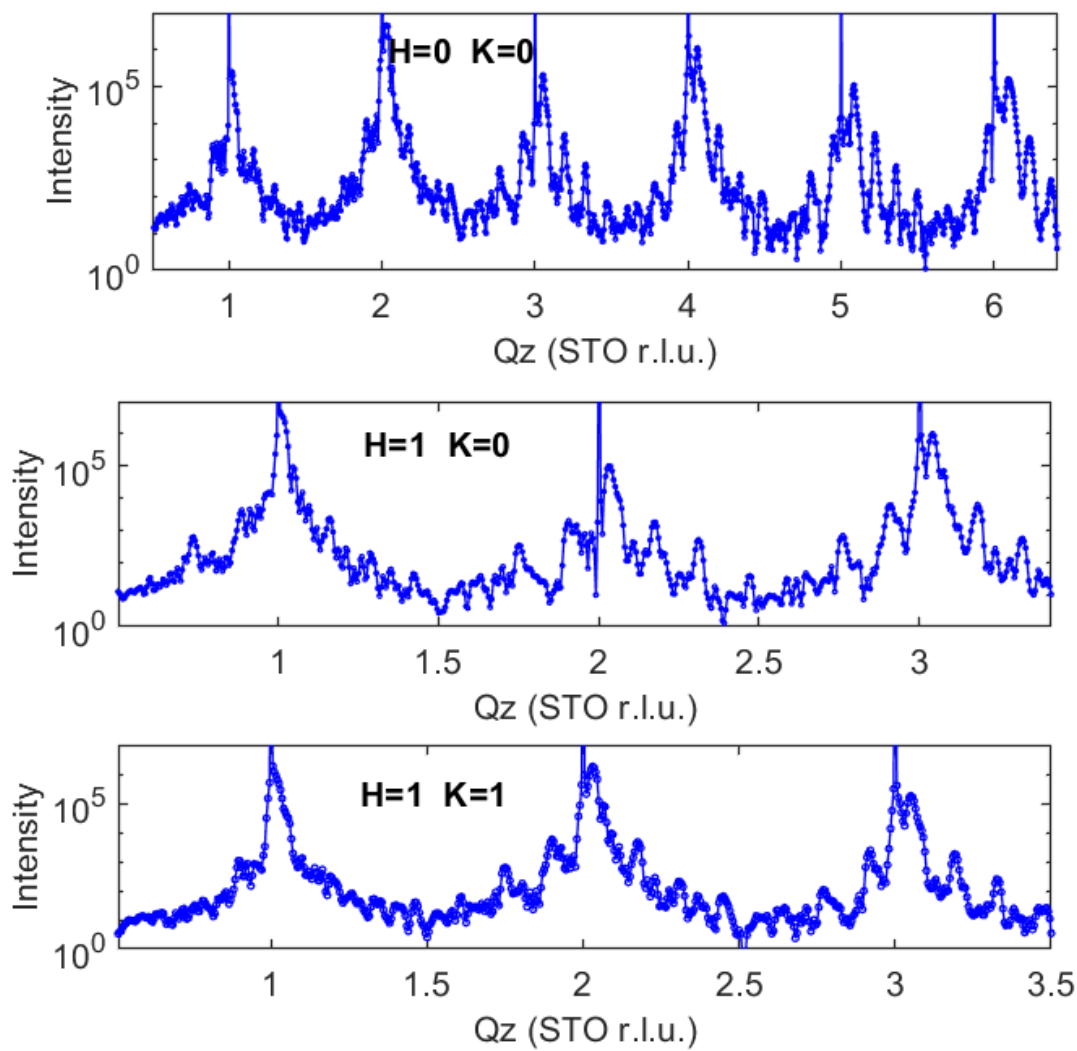

**Figure S7: Crystal truncation rods for a [LSCO (2)/LSMO (5)]x 6 superlattice grown on (001) SrTiO$_3$ by molecular beam epitaxy.** The diffraction pattern shows clear superlattice peaks, indicative of a uniformly repeating and coherently strained structure throughout the heterostructure.

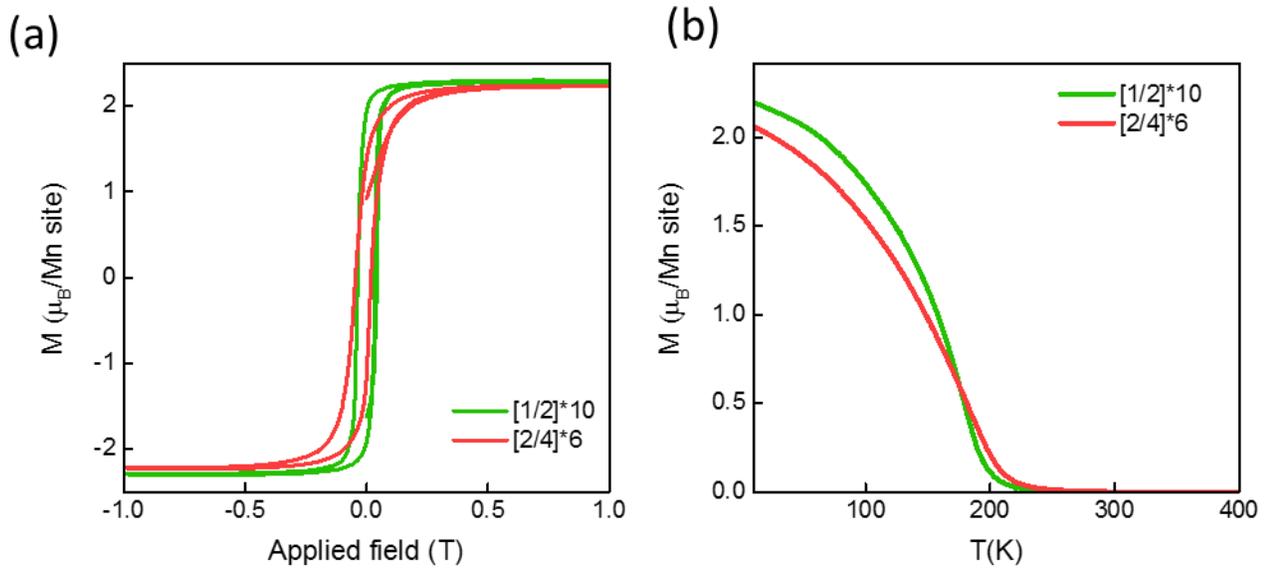

**Figure S8. Comparison of magnetization for LSCO/LSMO heterostructures with a 1:2 LSMO:LSCO ratio.** (a) Magnetization at 10 K as a function of applied in-plane magnetic field after zero field cool. (b) Magnetization as a function of temperature at for a [LSCO(1)/LSMO(2)] x10 and [LSCO(2)/LSMO(4)] x6 superlattice.

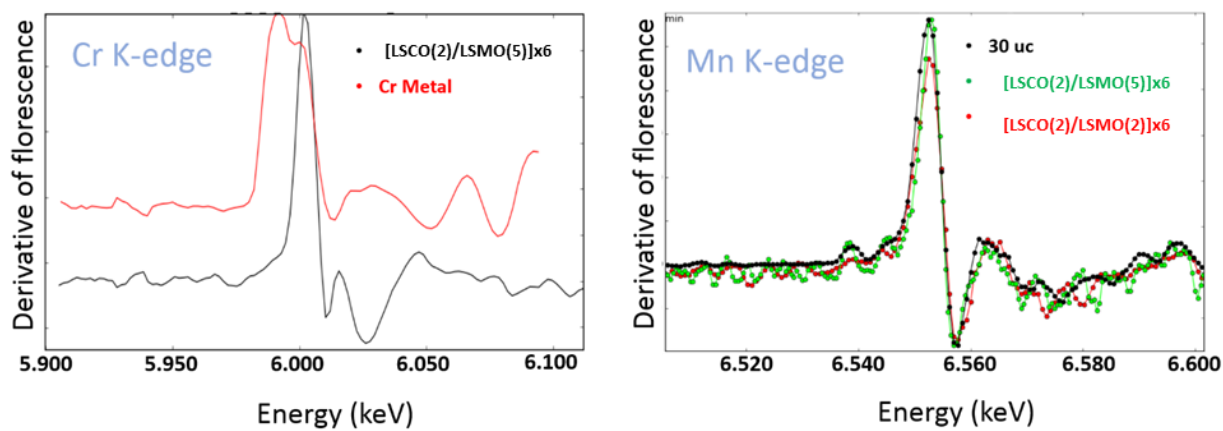

**Figure S9: Determination of the Cr and Mn valence states of the [LSCO(2)/LSMO (N)] superlattices by X-ray K-edge florescence spectroscopy.** The K-absorption edge is determined by the first inflection point of the florescence/absorption spectra. (left) Comparison of the first derivative of the florescence for a Cr metal and a [2 LSCO/5 LSMO]x6 superlattice. The absorption edges (first peak for each curve) occur at 5.9922 +/-0.0005 keV for Cr metal and 6.0035 +/-0.0005 keV for the superlattice. The energy shift of ~12 eV corresponds to a Cr valence of +3.3+/- 0.1 which is expected for $La_{0.7}Sr_{0.3}CrO_3$. (right panel) Comparison of the first derivative of the florescence for a 'bulk-like' 30 uc LSMO film, a [LSCO(2)/LSMO(5)]x6 superlattice and a [LSCO(2)/LSMO(2)]x10. The absorption edges for the 3 samples occurs at ~6.5526+/-0.0005 keV. The absence of a chemical shift indicates that the samples have identical oxidation states for the Mn ions.

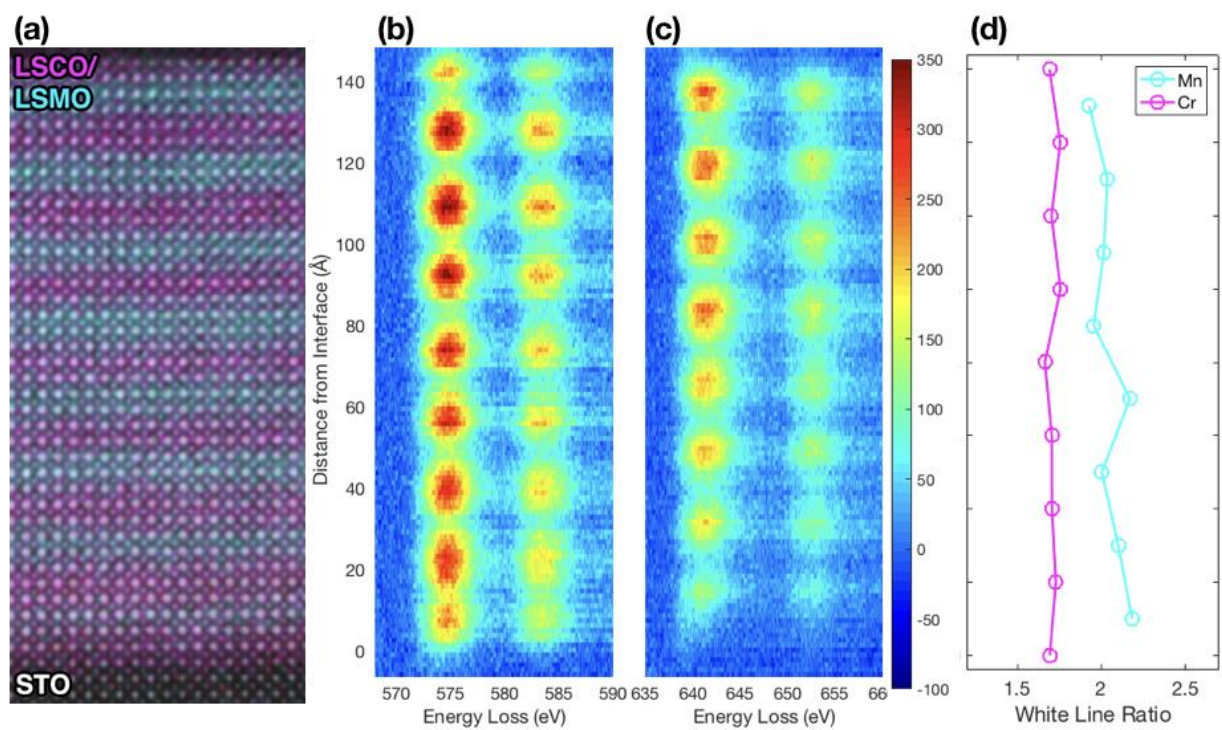

**Figure S10: Mn and Cr EELS edges as a function of film depth for [LSCO(2)/LSMO(3)]x8.** (a) HAADF STEM image of the heterostructure with the overlain EELS color map of alternating LSCO/LSMO layers. The energy shift and fine structure from averaged EEL spectra of Cr- (b) and Mn-containing layers (c). The intensity, denoted with the color bar, corresponds to the intensity of the peaks in arbitrary units. (d) The integrated WLR of the $L_3/L_2$-edge of Cr and Mn EEL spectra, indicative of the elements' valence. The shift, fine structure, and WLR remain mostly unchanged throughout the film's layers revealing a consistent valence.